\documentclass[%
 reprint,
superscriptaddress,
amsmath,amssymb,
aps,
floatfix,
]{revtex4-2}

\usepackage{graphicx}
\graphicspath{{./}{Manuscript/}}
\usepackage[utf8]{inputenc}
\DeclareUnicodeCharacter{2212}{\textminus}
\usepackage{bm}
\usepackage[colorlinks]{hyperref}

\usepackage{comment}
\usepackage{braket, placeins}
\usepackage{epsfig}

\newcommand{\micro}{\mu}
\newcommand{\SI}[2]{\ensuremath{#1\,\mathrm{#2}}}

\ifdefined\pdfadjustspacing
  \AtBeginDocument{%
    \pdfadjustspacing=2
    \begingroup
      \normalfont\normalsize\pdffontexpand\font 20 20 5 autoexpand\relax
      \small\pdffontexpand\font 20 20 5 autoexpand\relax
      \footnotesize\pdffontexpand\font 20 20 5 autoexpand\relax
    \endgroup
  }
\fi

\makeatletter
\let\original@frontmatter@footnote\frontmatter@footnote
\protected\def\separateauthornotes#1#2{%
  \original@frontmatter@footnote{#1}%
  \textsuperscript{,\,}%
  \original@frontmatter@footnote{#2}%
}
\long\def\frontmatter@footnote#1{%
  \begingroup
  \protected@edef\expanded@author@note{#1}%
  \expandafter\split@author@notes\expanded@author@note\@nil
  \endgroup
}
\long\def\split@author@notes#1#2\@nil{%
  \ifx#1\separateauthornotes
    #1#2%
  \else
    \original@frontmatter@footnote{#1#2}%
  \fi
}
\makeatother

\begin{document}

\preprint{APS/123-QED}

\title{Experimental Realization of a C-Shunt Flux-Qubit Coupler for Controlled-Z Gates between Transmons}

\def\affila{Beijing National Laboratory for Condensed Matter Physics, Institute of Physics, Chinese Academy of Sciences, Beijing 100190, China}
\def\affilb{School of Physics, University of Chinese Academy of Sciences, Beijing 100049, China}
\def\affilc{Beijing Academy of Quantum Information Sciences, Beijing 100193, China}
\def\affild{Hefei National Laboratory, Hefei 230088, China}
\def\affile{Songshan Lake Materials Laboratory, Dongguan 523808, Guangdong, China}

\newcommand{\fref}[1]{Fig.~\ref{#1}}
\newcommand{\eref}[1]{Eq.~(\ref{#1})}
\newcommand{\aref}[1]{Appendix~\ref{#1}}

\author{Weiping~Yuan}
\thanks{These authors contributed equally to this work.}
\affiliation{\affila}
\affiliation{\affilb}

\author{Silu~Zhao}
\thanks{These authors contributed equally to this work.}
\affiliation{\affila}

\author{Li~Li}
\thanks{\separateauthornotes{These authors contributed equally to this work.}{\href{mailto:lili@baqis.ac.cn}{lili@baqis.ac.cn}}}
\affiliation{\affila}
\affiliation{\affilc}

\author{Chenglin~Deng}
\affiliation{\affilc}

\author{Zhenghe~Liu}
\affiliation{\affila}
\affiliation{\affilb}

\author{Yu~Liu}
\affiliation{\affila}

\author{Tianming~Li}
\affiliation{\affila}
\affiliation{\affilb}

\author{Mingchuan~Wang}
\affiliation{\affila}
\affiliation{\affilb}

\author{Bingjie~Chen}
\affiliation{\affila}
\affiliation{\affilb}

\author{Guihan~Liang}
\affiliation{\affila}
\affiliation{\affilb}

\author{Jiacheng~Song}
\affiliation{\affila}
\affiliation{\affilb}

\author{Jinming~Guo}
\affiliation{\affila}
\affiliation{\affilb}

\author{Shuaili~Wang}
\affiliation{\affila}
\affiliation{\affilb}

\author{Mingxuan~Zhou}
\affiliation{\affila}
\affiliation{\affilb}

\author{Xiaohui~Song}
\affiliation{\affila}
\affiliation{\affild}

\author{Kai~Xu}
\email{kaixu@iphy.ac.cn}
\affiliation{\affila}
\affiliation{\affilc}
\affiliation{\affild}

\author{Heng~Fan}
\email{hfan@iphy.ac.cn}
\affiliation{\affila}
\affiliation{\affilb}
\affiliation{\affilc}
\affiliation{\affild}
\affiliation{\affile}

\author{Zhongcheng~Xiang}
\email{zcxiang@iphy.ac.cn}
\affiliation{\affila}
\affiliation{\affild}

\author{Dongning~Zheng}
\affiliation{\affila}
\affiliation{\affilb}
\affiliation{\affild}

\date{\today}

\begin{abstract}
Fast entangling gates and low-interaction idling require distinct operating conditions in superconducting quantum processors. We experimentally realize a transmon-C-shunt-transmon (TCT) device in which a flux-tunable, three-junction capacitively shunted flux-qubit coupler connects two transmon data qubits. The C-shunt coupler provides a flux-tunable, positively anharmonic spectrum for tailoring the interactions between the data qubits. Conditional Ramsey and swap measurements identify an idle region with near-zero residual ZZ interaction strength and suppressed transverse exchange. At this bias, simultaneous single-qubit interleaved randomized benchmarking yields $X$-gate fidelities above 99.7\%, consistent with individual operation. A \SI{45}{ns} nonadiabatic controlled-Z (CZ) gate achieves an interleaved randomized benchmarking fidelity estimate of 99.23(15)\%. Numerical simulations identify coupler energy relaxation as an important limitation on CZ-gate fidelity and indicate that extending the coupler lifetime can improve gate performance. These results demonstrate the experimental feasibility of C-shunt flux-qubit couplers for combining low-interaction idling with fast entangling operations between transmon qubits.
\end{abstract}
\maketitle

\frenchspacing
\tolerance=1000
\emergencystretch=1em


\section{\label{sec:introduction} INTRODUCTION}

Superconducting circuits have achieved single-qubit gate errors below $10^{-4}$~\cite{ZhiyuanLi2023_1QBtransmon9999,Bland2025_1QBtransmon9999} and two-qubit errors at or below $10^{-3}$~\cite{Marxer2026_999Device,Glaser2024_TTT9992QB}. The transmon qubit~\cite{Koch2007_Transmon} provides a practical circuit platform for quantum information processing. Fast reset~\cite{Zhou2021_FastResetTransmon} and high-fidelity readout~\cite{Spring2025_intrinsicPCfilterRO} have further supported the development of noisy intermediate-scale processors~\cite{Preskill2018_NISQ,YulinWu2021_Zhuchongzhi,Arute2019_Sycamore}. As these processors scale up, advances in cryogenic integration~\cite{Krinner2019_100scqubit} and quantum error correction~\cite{Acharya2023_Google_d=5surfacecode,Willow2025,Wang2025_qLDPC,lacroix2024_scalinglogiccolorcode} increase the need for precise interaction control.

Fast entangling gates require strong interactions, whereas low-error idling requires suppression of transverse exchange and residual ZZ interactions, quantified by $\zeta_{ZZ}$. A nonzero $\zeta_{ZZ}$ produces unwanted conditional phases during simultaneous operations~\cite{Mundada2019_ZZxtalkCancel} and can degrade cross-resonance gates~\cite{Zhao2023_CRgateZZmitigation}, whose unwanted Hamiltonian terms also require careful calibration~\cite{Sheldon2016_CRgateZZ}. A coupler must therefore provide distinct idle and gate operating points.

In transmon circuits, residual ZZ interactions arise from hybridization with non-computational levels~\cite{FeiYan2021_TTTCZgatetheory}. These interactions can be suppressed by coupling a transmon with negative anharmonicity~\cite{JaseungKu2020_UnwantedZZ}. Tunable transmon couplers use interference between direct and coupler-mediated exchange paths~\cite{FeiYan2018_TTTtheory,YuanXu2020_TTTCZgate,Collodo2020_TTT} to suppress idle interactions while enabling fast gates~\cite{Sung2021_TTT,Marxer2023_TTTCZgate,Stehlik2021_TTT}. Double-transmon couplers provide an alternative architecture for suppressing residual ZZ interactions while enabling high-fidelity CZ gates~\cite{RuiLi2024DTSCZGate,RuiLi2025_DTSPRA}. High-fidelity gate operation in these architectures requires accurate flux-pulse calibration~\cite{Li2024_TunableCouplerPulseCalibration}, careful frequency allocation to limit spectator interactions~\cite{Valles2026_FrequencyPositioning}, and control of leakage and phase errors~\cite{Zhang2026_CZCalibrationWorkflow}. The nearly harmonic spectrum of a transmon coupler constrains the freedom to optimize these requirements simultaneously.

Strongly anharmonic circuits offer additional spectral flexibility. Fluxonium supports long coherence~\cite{LongNguyen2019_FluxoniumReview,Somoroff2023_ms_fluxonium} and high-fidelity two-qubit gates~\cite{LeonDing2023_FTF}, while integer-fluxonium and bifluxon circuits offer partial protection against decoherence~\cite{Mencia2024_IntegerFluxonium,Ardati2024_IntegerFluxonium}. Fluxonium couplers also provide spectral flexibility~\cite{An2025_TFT}, but typically require Josephson-junction-array superinductors~\cite{LongNguyen2019_FluxoniumReview,Manucharyan2009_fluxonium}. A capacitively shunted flux qubit (C-shunt) provides a tunable spectrum with large positive anharmonicity without a superinductor~\cite{Yan2016_CshuntFluxQubit,Yurtalan2021_CshuntFluxCircuit}. Its use as a coupler for suppressing residual transmon interactions has been proposed theoretically~\cite{Heunisch2023_CSFQcplrtheory}.

Existing experiments include two-qubit control with a C-shunt computational qubit coupled to a transmon~\cite{JaseungKu2020_UnwantedZZ} and a capacitively shunted double-transmon coupler with transmon-like and C-shunt-flux-qubit-like modes~\cite{RuiLi2025_DTSPRA}. Here, we realize a three-junction C-shunt flux-qubit coupler between two transmon data qubits.

Our transmon-C-shunt-transmon (TCT) device exhibits suppressed exchange and near-zero $\zeta_{ZZ}$ at idle, with comparable simultaneous and individual single-qubit $X$-gate fidelities. A \SI{45}{ns} nonadiabatic CZ gate achieves an interleaved randomized benchmarking fidelity estimate of 99.23(15)\%. Numerical analysis identifies the short energy-relaxation time of the coupler as an important source of CZ-gate error. With the measured data-qubit lifetimes and a fixed optimized pulse, relaxation-only simulations predict a CZ-gate fidelity approaching 99.8\% when the coupler lifetime is increased to \SI{5}{\micro s}. These results motivate further improvements in coupler coherence through device design and fabrication optimization. Finally, we discuss the scalability of the TCT architecture through numerical simulations.

\section{\label{sec:principle_of_tft} TCT ARCHITECTURE AND OPERATION}

\begin{figure}[t]
    \centering
    \includegraphics[width=\columnwidth]{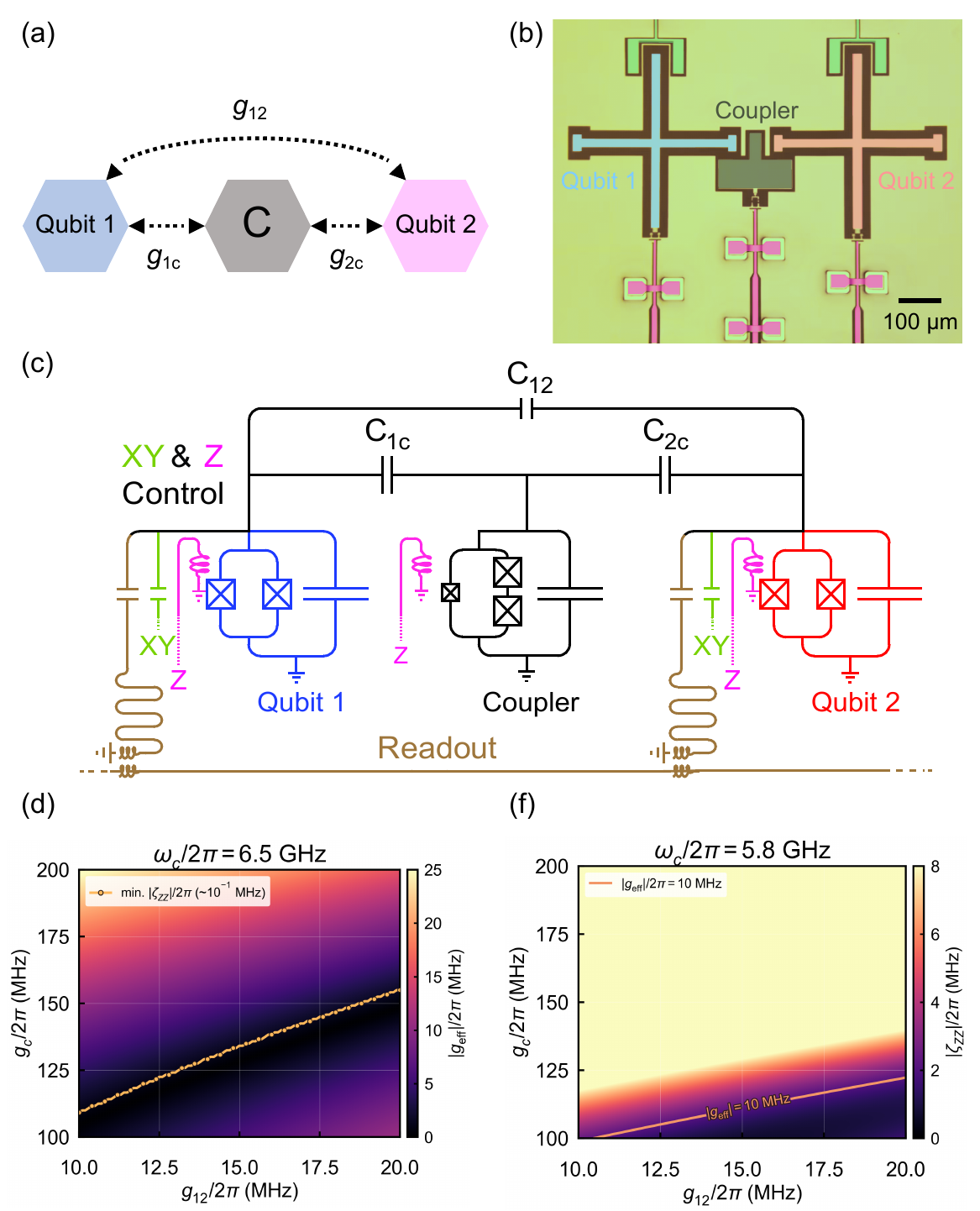}
    \vspace{0pt}
    \caption{TCT architecture and numerical design window. (a) Schematic connectivity of Qubit~1, the C-shunt coupler, and Qubit~2. (b) False-color device micrograph. (c) Equivalent circuit of the TCT device, showing Qubit~1 and Qubit~2, the central C-shunt flux-qubit coupler, the qubit-coupler capacitances $C_{1c}$ and $C_{2c}$, the residual direct capacitance $C_{12}$, and the control and readout lines. (d) Numerical $|g_{\mathrm{eff}}|/2\pi$ map for $\omega_c/2\pi=\SI{6.5}{GHz}$, with the trajectory minimizing $|\zeta_{ZZ}|$ overlaid. (f) Numerical $|\zeta_{ZZ}|/2\pi$ map for $\omega_c/2\pi=\SI{5.8}{GHz}$, with the $|g_{\mathrm{eff}}|/2\pi=\SI{10}{MHz}$ contour overlaid.}
    \label{fig:tft_schematics}
    \vspace{-1.0em}
\end{figure}

The TCT architecture consists of two transmon data qubits capacitively coupled through a C-shunt flux-qubit coupler. Figure~\ref{fig:tft_schematics}(a,b) shows the connectivity and device micrograph. The equivalent circuit in \fref{fig:tft_schematics}(c) contains the qubit-coupler capacitances $C_{1c}$ and $C_{2c}$ and the smaller direct qubit-qubit capacitance $C_{12}$. Logical states are encoded in the two lowest levels of each data qubit, and external flux controls the coupler spectrum.

The coupler uses a three-junction flux-qubit loop~\cite{Orlando1999_FluxQubit} with a large shunt capacitance~\cite{Yan2016_CshuntFluxQubit,You2007_LowDecoherenceFluxQubit,Steffen2010_HybridFluxQubit}. Its shunt capacitance suppresses charge dispersion, while the loop retains magnetic-flux tunability, as established in C-shunt experiments~\cite{Yan2016_CshuntFluxQubit,Yurtalan2021_CshuntFluxCircuit,Abdurakhimov2019_LongLivedCshuntFluxQubit}. In the present device, this combination produces a flux-tunable mode with a large positive anharmonicity near idle. The fabrication process and design parameters are detailed in Appendix~\ref{appendix:fabrication}. The measured device parameters are presented below, and the circuit model is derived in Appendix~\ref{app:tct_circuit_hamiltonian}.

To analyze the idle and gate operating points, we describe the TCT device, to lowest order, as two weakly anharmonic modes, Qubit~1 and Qubit~2, capacitively coupled through a tunable nonlinear C-shunt mode. The high-frequency internal coupler mode is frozen, leaving one active coupler mode. In the rotating-wave approximation, the effective Hamiltonian is
\begin{align} \label{eqn:tft_hamiltonian}
\begin{split}
    \hat{H}/\hbar = & \sum_{\mu=1,c,2}
    \left[\omega_{\mu}(\Phi_{c}) \hat{a}_{\mu}^{\dagger}\hat{a}_{\mu}
    + \frac{\alpha_{\mu}(\Phi_{c})}{2}
    \hat{a}_{\mu}^{\dagger}\hat{a}_{\mu}^{\dagger}\hat{a}_{\mu}\hat{a}_{\mu}\right] \\
    & + g_{1c}\left(\hat{a}_{1}^{\dagger}\hat{a}_{c}+\hat{a}_{1}\hat{a}_{c}^{\dagger}\right)
    + g_{2c}\left(\hat{a}_{2}^{\dagger}\hat{a}_{c}+\hat{a}_{2}\hat{a}_{c}^{\dagger}\right) \\
    & + g_{12}\left(\hat{a}_{1}^{\dagger}\hat{a}_{2}+\hat{a}_{1}\hat{a}_{2}^{\dagger}\right).
\end{split}
\end{align}
Here $\hat a_\mu$ annihilates an excitation in mode $\mu=1,c,2$, and $\omega_\mu$, $\alpha_\mu$, and $g_{\mu\nu}$ are angular frequencies. The externally applied coupler flux $\Phi_c$ tunes $\omega_c$ and $\alpha_c$, while the data-qubit frequencies remain nearly fixed. The capacitive qubit-coupler couplings $g_{1c}$ and $g_{2c}$ may also depend on flux, although this dependence is not shown explicitly in the notation. The bare qubit-qubit exchange $g_{12}$ is defined by the full capacitance matrix.

To obtain the effective transverse exchange, we eliminate the coupler to second order using a Schrieffer--Wolff transformation. Retaining the leading rotating-wave terms yields
\begin{align}
    g_{\mathrm{eff}} \approx g_{12}
    + \frac{g_{1c}g_{2c}}{2}
    \left(\frac{1}{\Delta_{1c}}+\frac{1}{\Delta_{2c}}\right),
\end{align}
where $\Delta_{ic}=\omega_i-\omega_c$. The expression combines the direct exchange $g_{12}$ with a virtual exchange amplitude controlled by the coupler detuning. Their cancellation defines the exchange-off bias.

The residual ZZ interaction strength $\zeta_{ZZ}$ is defined by the conditional shift of the dressed eigenenergies:
\begin{align} \label{eqn:zz_definition_cplr}
    \zeta_{ZZ} = (E_{101} - E_{100} - E_{001} + E_{000})/\hbar,
\end{align}
where the subscripts denote the excitation numbers of Qubit~1, the coupler, and Qubit~2. Here $\zeta_{ZZ}$ is an angular frequency, and $\zeta_{ZZ}/2\pi$ is reported in frequency units. Thus the two-qubit logical state $|11\rangle$ is the dressed state connected to $|101\rangle$ when the coupler remains in its ground state. The perturbative expression for $\zeta_{ZZ}$ contains three leading contributions,
\begin{align}
    \zeta_{ZZ}\simeq
    \zeta_{ZZ}^{(2)}
    +\zeta_{ZZ}^{(3)}
    +\zeta_{ZZ}^{(4)}.
\end{align}
The second-order term scales as $g_{12}^{2}$ and depends on the anharmonicities of Qubit~1 and Qubit~2. The third-order term, proportional to $g_{12}g_{1c}g_{2c}$, describes interference between the direct qubit-qubit path and the coupler-mediated path. The fourth-order term, proportional to $g_{1c}^{2}g_{2c}^{2}$, is the purely coupler-mediated nonlinear contribution and contains denominators involving the coupler anharmonicity $\alpha_c$. Thus $\zeta_{ZZ}$ depends on the energy shifts induced by hybridization with $|200\rangle$, $|002\rangle$, and coupler-excited two-excitation states, beyond the single-excitation processes that determine $g_{\mathrm{eff}}$. The counter-rotating corrections to the effective exchange and the perturbative expressions for the residual ZZ interaction are given in Appendix~\ref{app:tct_circuit_hamiltonian}.

Exchange cancellation and low-$\zeta_{ZZ}$ idling therefore impose distinct conditions on the same spectrum. The C-shunt anharmonicity provides an additional parameter for reconciling these conditions while retaining a gate bias with stronger interactions.

The design maps in \fref{fig:tft_schematics}(d,f) illustrate this spectral-engineering window for parameters representative of the measured device. At $\omega_c/2\pi=\SI{5.8}{GHz}$, the $|g_{\mathrm{eff}}|/2\pi\simeq\SI{10}{MHz}$ contour passes through a region where $|\zeta_{ZZ}|/2\pi$ remains only a few megahertz [\fref{fig:tft_schematics}(f)]. At the higher coupler frequency $\omega_c/2\pi=\SI{6.5}{GHz}$, a region of small $|\zeta_{ZZ}|$ appears near exchange cancellation, with a numerical minimum of $|\zeta_{ZZ}|/2\pi\simeq\SI{56}{kHz}$ [\fref{fig:tft_schematics}(d)]. These maps guide the selection of the idle and gate biases characterized below.

\section{\label{sec:experiment_result} Device Spectroscopy and Idle-Point Characterization}

\begin{figure*}[t]
    \centering
    \includegraphics[width=0.94\textwidth]{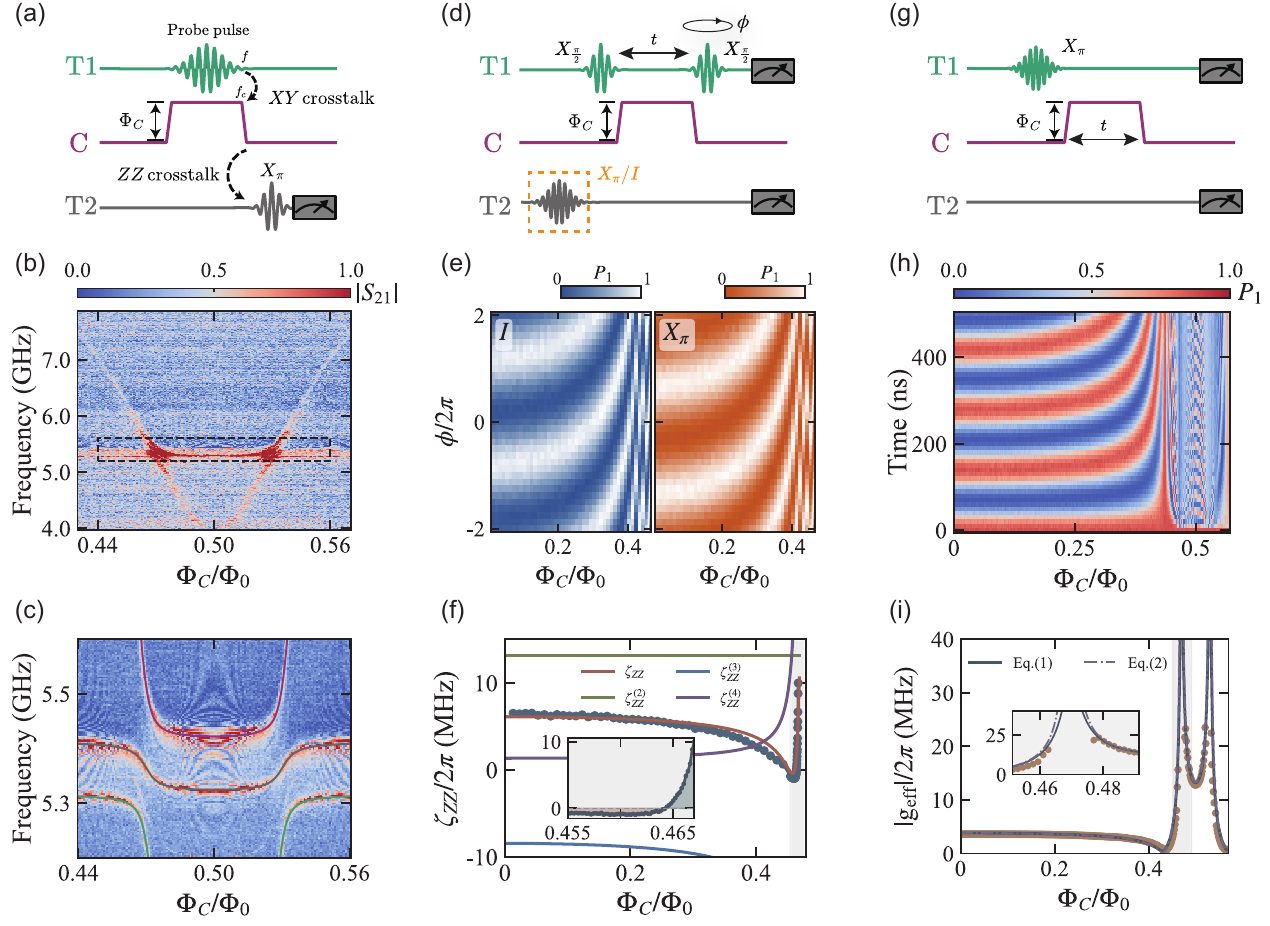}
    \vspace{-1.1em}
    \caption{Basic characterization of the TCT device. (a) Pulsed spectroscopy sequence: an XY probe drives Qubit~1, and the coupler-dependent response is detected through Qubit~2. (b) Broad C-shunt coupler spectroscopy versus external flux. (c) Enlarged spectroscopy near the data-qubit transitions, with overlaid three-mode transition curves. (d) Pulsed conditional Ramsey sequence for measuring the residual ZZ interaction, with the spectator qubit prepared using either $I$ or $X_\pi$. (e) Corresponding Ramsey maps. (f) Extracted $\zeta_{ZZ}/2\pi$ and theoretical contributions. (g) Pulse sequence for extracting the effective transverse coupling from swap oscillations. (h) Swap-oscillation data. (i) Extracted $|g_{\mathrm{eff}}|/2\pi$ and theoretical curves. Insets in (f,i) enlarge the operating window.}
    \label{fig:zz_vs_cplrflux}
    \vspace{-0.9em}
\end{figure*}

\subsection{Device Parameters and Spectroscopy}

We characterize the device in Fig.~\ref{fig:tft_schematics} using spectroscopy, conditional Ramsey measurements, and swap oscillations.

Table~\ref{tab:tft_param} summarizes the device transition frequencies, anharmonicities, energy-relaxation times, and coupling strengths. The tabulated $f_{10}$ values are \SI{5.334}{GHz} for Qubit~1, \SI{5.43}{GHz} for Qubit~2, and \SI{4.03}{GHz} for the C-shunt coupler. Qubit~1 and Qubit~2 have measured anharmonicities of approximately \SI{-256}{MHz} and \SI{-254}{MHz}, whereas the coupler value $\alpha_c/2\pi\simeq\SI{703}{MHz}$ is derived from the circuit model fitted to the measured spectra (Appendix~\ref{app:tct_device_spectroscopy}). The energy-relaxation times $T_1$ listed in the same table are \SI{17.50}{\micro s}, \SI{25.17}{\micro s}, and \SI{1.1}{\micro s}, respectively. The shorter coupler lifetime motivates the relaxation analysis in Sec.~\ref{sec:gate_error_analysis}.

The pulsed spectra in \fref{fig:zz_vs_cplrflux}(b,c) resolve the flux-tunable coupler-like branch and its avoided crossings with the two data-qubit transitions. The minimum coupler frequency and coupling strengths are extracted by fitting these experimental spectra using the equations in \aref{app:tct_device_spectroscopy}. The fit gives a minimum coupler frequency near \SI{4.00}{GHz}, qubit-coupler coupling strengths of $g_{1c}/2\pi\simeq\SI{101}{MHz}$ and $g_{2c}/2\pi\simeq\SI{105}{MHz}$, and a direct qubit-qubit coupling strength of $g_{12}/2\pi\simeq\SI{7.5}{MHz}$. The coupling strengths are also listed in Table~\ref{tab:tft_param}.

\begin{table}[t]
    \centering
    \caption{Summary of device parameters.}
    \label{tab:tft_param}
    \small
    \renewcommand{\arraystretch}{1.15}
    \setlength{\tabcolsep}{3pt}
    \begin{tabular*}{\columnwidth}{@{\extracolsep{\fill}}p{0.51\columnwidth}ccc@{}}
        \hline
        \hline
        Quantity, symbol (unit) & Qubit~1 & Qubit~2 & Coupler \\
        \hline
        \raggedright Transition frequency,\newline $f_{10}$ (GHz) & $5.334$ & $5.43$ & $4.03$ \\
        Anharmonicity, $\alpha_i/2\pi$ (MHz) & $-256$ & $-254$ & $703$ \\
        \raggedright Energy relaxation time,\newline $T_1$ ($\mu$s) & $17.50$ & $25.17$ & $1.1$ \\
        \raggedright Qubit-coupler coupling,\newline $g_{ic}/2\pi$ (MHz) & $101$ & $105$ & --- \\
        \raggedright Direct qubit-qubit coupling,\newline $g_{12}/2\pi$ (MHz) & \multicolumn{2}{c}{$7.5$} & --- \\
        \hline
        \hline
    \end{tabular*}
\end{table}

\vspace{-0.6em}
\subsection{Residual ZZ Cancellation and Tunable Exchange Coupling}

We next characterized the two interactions that determine the idle and gate operating points. The residual ZZ interaction was measured with the pulsed conditional Ramsey sequence in \fref{fig:zz_vs_cplrflux}(d). At each coupler bias, a square flux pulse was inserted between two Ramsey $\pi/2$ pulses on one data qubit, while the other data qubit was prepared in either $|0\rangle$ or $|1\rangle$. Representative Ramsey maps are shown in \fref{fig:zz_vs_cplrflux}(e). Fitting each fringe to a sinusoidal response gives the conditional phase difference
\begin{equation}
    \Delta\phi = \phi_{|1\rangle}-\phi_{|0\rangle},
\end{equation}
and hence the residual ZZ interaction strength during the flux pulse,
\begin{equation}
    \frac{\zeta_{ZZ}}{2\pi}=\frac{\Delta\phi}{2\pi\tau_{\mathrm{eff}}},
\end{equation}
where $\tau_{\mathrm{eff}}$ is the effective conditional-phase accumulation time used in the calibration.

The extracted $\zeta_{ZZ}/2\pi$ in \fref{fig:zz_vs_cplrflux}(f) changes sign as the coupler flux is tuned. The broad scan shows that $\zeta_{ZZ}/2\pi$ can be varied from approximately \SI{-5.2}{MHz} to \SI{10.5}{MHz}, while the inset resolves the zero crossing near $\Phi_{\mathrm{ext}}/\Phi_{0}\simeq0.46$. This point is used as the low-crosstalk idle region, where the direct and coupler-mediated contributions to the conditional frequency shift cancel.

The effective transverse coupling was measured independently with the swap sequence in \fref{fig:zz_vs_cplrflux}(g). A single excitation was prepared on one data qubit, the coupler was biased to a selected flux, and the population exchange between the two data qubits was recorded as a function of interaction time. Fitting the chevron data in \fref{fig:zz_vs_cplrflux}(h) yields the effective coupling $g_{\mathrm{eff}}/2\pi$ shown in \fref{fig:zz_vs_cplrflux}(i). The coupling is strongly suppressed near the idle bias and increases rapidly when the coupler is pulsed toward the gate region, reaching the \SI{10}{MHz} scale in the enlarged operating window.

Together, the Ramsey and swap measurements locate the operating biases used for single-qubit control and CZ gates in the following section.

\begin{figure*}[t]
    \centering
    \includegraphics[width=\textwidth]{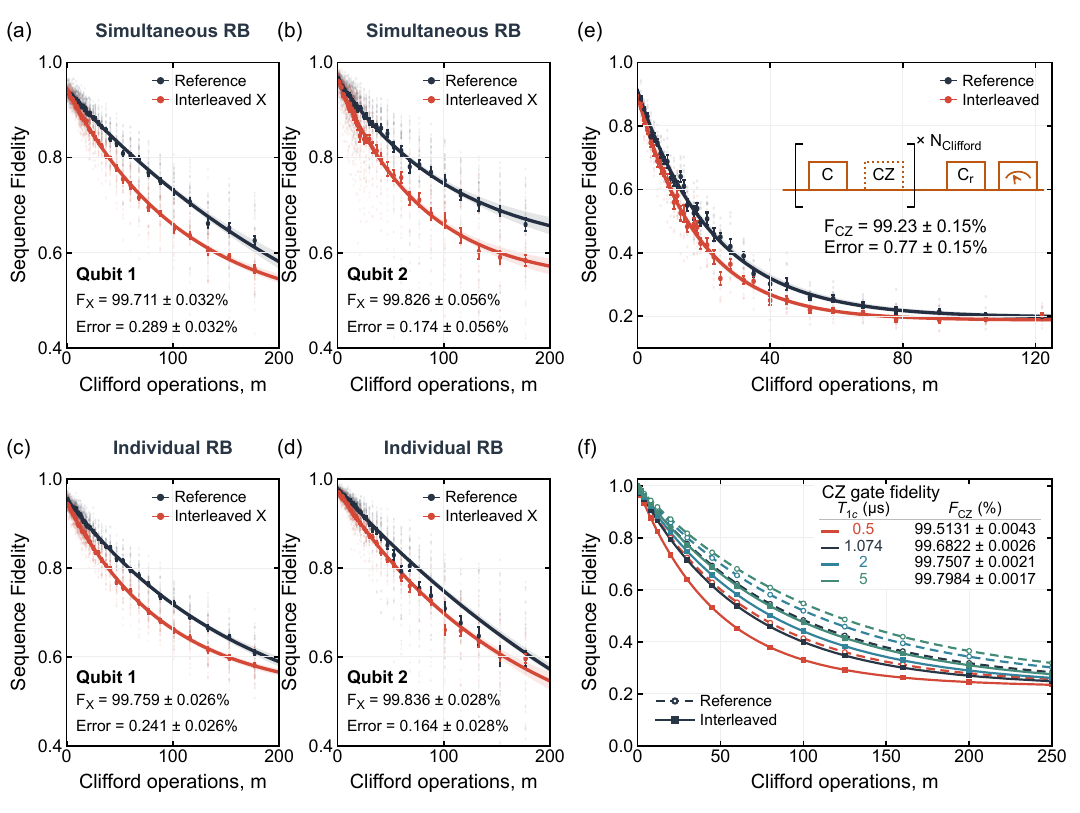}
    \caption{Single-qubit and CZ randomized benchmarking (RB) and error analysis.
    (a,b) Simultaneous $X$-interleaved RB for Qubit~1 and Qubit~2. (c,d) Corresponding individual measurements.
    (e) Reference and CZ-interleaved RB for the \SI{45}{ns} CZ gate, giving $F_{\mathrm{CZ}}=99.23(15)\%$. The inset shows the RB sequence.
    In (a)--(e), the reference and interleaved data are shown in dark colors and red, respectively.
    (f) Simulated reference (dashed lines, open circles) and CZ-interleaved (solid lines, squares) RB for four assumed coupler lifetimes, with data-qubit $T_1=17.50$ and \SI{25.17}{\micro s} from Table~\ref{tab:tft_param}. The simulations use a fixed, numerically optimized \SI{45}{ns} CZ pulse and \SI{20}{ns} single-qubit rotations, with relaxation as the only noise source.}
    \label{fig:cz_rb}
\end{figure*}

\section{Gate Calibration and Benchmarking}

\subsection{Single-Qubit Gate Benchmarking}

We calibrated the microwave single-qubit gates at the idle bias identified above. Each primitive single-qubit rotation used a \SI{20}{ns} Gaussian microwave pulse. We calibrated the amplitude, detuning, quadrature, and virtual-$Z$ corrections, including a DRAG quadrature correction to reduce leakage from the weakly anharmonic data qubits~\cite{Motzoi2009_DRAGpulse}. The calibrated control set generates the single-qubit Clifford group from the microwave rotations $\{I,\pm X_{\pi},\pm Y_{\pi},\pm X_{\pi/2},\pm Y_{\pi/2}\}$.

We characterized the calibrated $X$ gates using reference and interleaved randomized benchmarking, both with simultaneous control of the two qubits and with each qubit driven individually~\cite{Knill2008_CliffordRB,Magesan2012_CliffordRB,Gambetta2012_Sim1QBRBandZZ,Magesan2012_InterleavedRB}. The measured return probability is fitted to $A p^m+B$, where $m$ is the number of random Clifford operations. In the interleaved sequences, an additional $X$ gate follows each Clifford. For a single qubit, the decay-parameter ratio gives $F_X=1-(1-p_{\mathrm{int}}/p_{\mathrm{ref}})/2$.

The extracted $X$-gate fidelities under simultaneous operation are $F_X=99.711(32)\%$ for Qubit~1 and $99.826(56)\%$ for Qubit~2 [\fref{fig:cz_rb}(a,b)]. The corresponding individual measurements yield $99.759(26)\%$ and $99.836(28)\%$, respectively [\fref{fig:cz_rb}(c,d)], where the values in parentheses indicate uncertainties of one statistical standard deviation. The simultaneous errors are higher by 0.048 and 0.010 percentage points, respectively, but neither difference is resolved within the statistical uncertainty. This agreement supports low-crosstalk single-qubit control at the idle point.

\subsection{CZ Gate Implementation, Calibration, and Benchmarking}

The TCT device implements a fast nonadiabatic CZ gate by pulsing the C-shunt coupler
from the idle point at $\Phi_c/\Phi_0\simeq0.46$, where exchange is suppressed and $\zeta_{ZZ}\simeq0$, to a nearby
high-interaction bias and back. During this excursion, the dressed $|11\rangle$
state hybridizes with higher non-computational states and accumulates a conditional
phase. The pulse must therefore satisfy two coupled requirements: the conditional
phase must reach $\pi$, and the population must return to the computational
subspace at the end of the gate, as in nonadiabatic CZ protocols~\cite{Sung2021_TTT,Negirneac2021_CZ}.

In the ideal limit of complete population return, the operation before
single-qubit phase compensation is
\begin{equation}
U_{\phi}=
\begin{pmatrix}
1 & 0 & 0 & 0 \\
0 & e^{i\phi_{01}} & 0 & 0 \\
0 & 0 & e^{i\phi_{10}} & 0 \\
0 & 0 & 0 & e^{i(\phi_{01}+\phi_{10}+\phi_{ZZ})}
\end{pmatrix}.
\end{equation}
Here $\phi_{ZZ}$ is the accumulated conditional phase, distinct from the residual ZZ interaction strength $\zeta_{ZZ}$. Setting $\phi_{ZZ}=\pi$ and removing the single-qubit phases $\phi_{01}$ and
$\phi_{10}$ with virtual $Z$ rotations gives
$U_{\mathrm{CZ}}=\mathrm{diag}(1,1,1,-1)$.

To satisfy these conditions in our device, we jointly tune the qubit and coupler flux-pulse amplitudes and the interaction duration. Population-return and conditional-phase scans identify an operating region that combines return to the computational subspace with a phase near $\pi$. Virtual-$Z$ corrections compensate for the single-qubit dynamical phases. The calibrated CZ gate is implemented using a \SI{45}{ns} square flux pulse.

The reference and CZ-interleaved RB data in \fref{fig:cz_rb}(e) are fitted using~\cite{Magesan2012_InterleavedRB}
\begin{equation}
\begin{aligned}
    F_{\mathrm{seq}}(m)&=A p^m+B,\\[3pt]
    r_{\mathrm{CZ}}&=\frac{3}{4}\left(1-\frac{p_{\mathrm{int}}}{p_{\mathrm{ref}}}\right).
\end{aligned}
    \label{eqn:cz_irb_error}
\end{equation}
The factor $3/4$ follows from the dimension $d=4$ of the two-qubit computational space. Each decay has independently fitted parameters $A$ and $B$. The CZ fidelity is $F_{\mathrm{CZ}}=1-r_{\mathrm{CZ}}$. The fits give $p_{\mathrm{ref}}=0.9570(11)$ and $p_{\mathrm{int}}=0.9472(16)$, with $p_g\equiv p_{\mathrm{int}}/p_{\mathrm{ref}}=0.9897(21)$. Equation~\eqref{eqn:cz_irb_error} yields $r_{\mathrm{CZ}}=0.771(154)\%$ and $F_{\mathrm{CZ}}=99.23(15)\%$ under the standard IRB assumptions~\cite{Magesan2012_InterleavedRB}.

\subsection{Gate Error Analysis}
\label{sec:gate_error_analysis}

We analyze the benchmarking results by estimating the single-qubit relaxation contribution and modeling the dependence of CZ fidelity on coupler lifetime. For the nonadiabatic CZ gate, transient hybridization with the coupler can produce irreversible relaxation even when coherent evolution returns population to the computational subspace at the end of the pulse~\cite{Sung2021_TTT,Abad2025_OutsideSubspace}.

Simultaneous benchmarking yields $X$-gate errors of 0.289(32)\% for Qubit~1 and 0.174(56)\% for Qubit~2. For a rotation of duration $t_{\mathrm{1Q}}=\SI{20}{ns}$, the leading contribution of energy relaxation to the average gate infidelity~\cite{Pedersen2007_incoherentlimit} is given by~\cite{Abad2022_WeakDissipation}
\begin{equation}
    r_{\mathrm{1Q}}^{T_1}\simeq\frac{t_{\mathrm{1Q}}}{3T_1}.
    \label{eqn:tct_1qb_t1_limit}
\end{equation}
Substituting the measured idle-point lifetimes $T_{1,1}=\SI{17.50}{\micro s}$ and $T_{1,2}=\SI{25.17}{\micro s}$ from Table~\ref{tab:tft_param} into Eq.~\eqref{eqn:tct_1qb_t1_limit} gives relaxation errors of 0.038\% and 0.026\%, respectively. These estimates account for approximately 13\% and 15\% of the measured simultaneous gate errors. Pulse-shape optimization, more accurate calibration, and suppression of dephasing and drive crosstalk provide complementary routes to improving single-qubit control.

We next assess the contribution of coupler relaxation to the measured CZ error of $0.771(154)\%$ by simulating the gate dynamics. We use a three-mode Hamiltonian with four levels per mode and counter-rotating capacitive interactions, and represent relaxation by a Lindblad equation~\cite{Sung2021_TTT}:
\begin{equation}
    \dot\rho=-\frac{i}{\hbar}[H(t),\rho]
    +\sum_{j=1,c,2}\frac{\mathcal D[a_j]\rho}{T_{1,j}},
    \label{eqn:cz_lindblad_t1}
\end{equation}
where $a_j$ is the lowering operator of mode $j$ and $\mathcal D[a]\rho=a\rho a^\dagger-\{a^\dagger a,\rho\}/2$. The commutator describes coherent evolution under the control pulse, and the dissipative terms describe energy relaxation of the qubits and coupler. We fix the data-qubit lifetimes to the values in Table~\ref{tab:tft_param} and vary $T_{1,c}$ from \SI{0.5}{\micro s} to \SI{5}{\micro s}, using the same numerically optimized \SI{45}{ns} pulse. Solving Eq.~\eqref{eqn:cz_lindblad_t1} gives a noisy CZ channel for each assumed coupler lifetime. These channels are then used in the reference and interleaved RB simulations.

The simulated reference and interleaved sequences include relaxation during CZ gates, single-qubit rotations, and waiting periods. Following the distinction between leakage and seepage in leakage RB~\cite{Wood2018_LeakageRB}, we adopt the absorbing-loss limit, with no return from the leakage subspace. We fit the computational-subspace survival probability $S_b$ and the return probability $P_{00,b}$ to the following forms, where $b$ labels the reference ($\mathrm{ref}$) or CZ-interleaved ($\mathrm{int}$) ensemble:
\begin{equation}
\begin{aligned}
    S_b(m)&=B_b s_b^m,\\
    P_{00,b}(m)-S_b(m)/4&=A_b p_b^m,\\[2pt]
    F_{\mathrm{CZ}}^{\mathrm{sim}}&\simeq\frac{1}{4}
    \left(\frac{s_{\mathrm{int}}}{s_{\mathrm{ref}}}
    +3\frac{p_{\mathrm{int}}}{p_{\mathrm{ref}}}\right).
\end{aligned}
    \label{eqn:cz_simulated_rb}
\end{equation}
For a twirled absorbing-loss channel in $d=4$, the average fidelity is $(s+3p)/4$~\cite{Wood2018_LeakageRB}. The use of interleaved ratios additionally assumes gate-independent noise with factorized decay parameters~\cite{Magesan2012_InterleavedRB}. For nonzero seepage, the more general leakage-RB fit is required. In the absence of leakage, the expression reduces to the conventional IRB estimator.

\begin{figure*}[t]
    \centering
    \includegraphics[width=\textwidth]{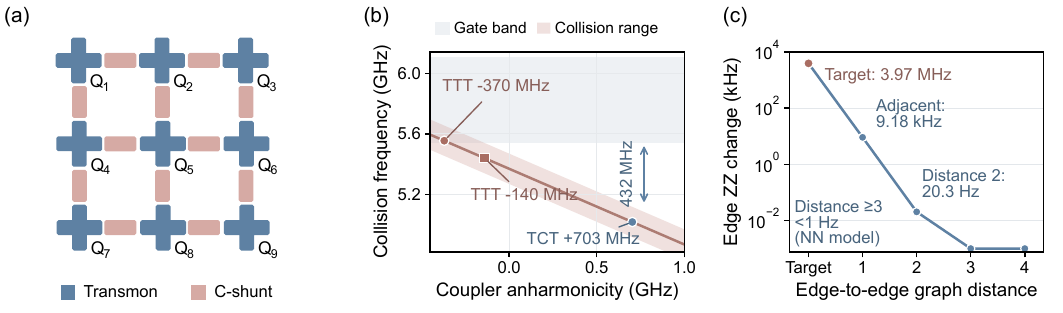}
    \caption{Numerical analysis of spectral collisions and spectator interactions. (a) Schematic of a possible $3\times3$ nearest-neighbor array of negatively anharmonic transmon data qubits $Q_1$--$Q_9$ (blue crosses), linked by positively anharmonic, flux-tunable C-shunt couplers (pink links), denoted $C_{ij}$ for the link between $Q_i$ and $Q_j$. (b) Bare-state $|020\rangle$--$|101\rangle$ collision frequency for a representative cell $Q_1$--$C_{12}$--$Q_2$, versus coupler anharmonicity. Other cells are omitted from this single-cell estimate. The pale pink band shows the collision range over the fixed mean-qubit-frequency allocation, and the gray band denotes the gate-control band. Markers identify representative TTT values and the TCT device-model value. (c) Gate-induced change in the residual ZZ interaction strength on each link, plotted as $|\Delta\zeta_{ZZ}|/2\pi$, for a six-qubit chain, mapped onto the path $Q_1$--$Q_2$--$Q_3$--$Q_6$--$Q_5$--$Q_4$ in (a). Only $C_{12}$ is activated, while $C_{23}$, $C_{36}$, $C_{56}$, and $C_{45}$ remain at idle. These five links correspond to distances 0--4 along the retained chain. All other array links are excluded from the model. Values below \SI{1}{Hz} are shown at the plotting floor.}
    \label{fig:tct_scalability}
    \vspace{-0.7em}
\end{figure*}

Figure~\ref{fig:cz_rb}(f) shows the simulated RB decays and extracted CZ fidelities. Increasing $T_{1,c}$ from \SI{0.5}{\micro s} to \SI{5}{\micro s} raises the simulated fidelity from $99.5131(43)\%$ to $99.7984(17)\%$. At \SI{1.074}{\micro s} and \SI{2}{\micro s}, the fidelities are $99.6822(26)\%$ and $99.7507(21)\%$, respectively. Both the reference and interleaved RB decays become slower as $T_{1,c}$ increases because the reference Clifford sequences also contain native CZ gates.

Increasing the coupler lifetime from \SI{2}{\micro s} to \SI{5}{\micro s} improves the simulated CZ fidelity by 0.048 percentage points. The diminishing improvement reflects the increasing relative importance of data-qubit relaxation and residual errors for the fixed pulse. At the representative lifetime of \SI{1.074}{\micro s}, the simulated CZ error is $0.3178(26)\%$, compared with the measured $0.771(154)\%$.

The simulated dependence on coupler lifetime points to improved coupler coherence as one route to higher CZ fidelity. The difference between the simulated and measured errors also motivates improvements in gate control. These include suppression of leakage and residual conditional-phase errors~\cite{Sung2021_TTT,Negirneac2021_CZ} and correction of flux-pulse distortion~\cite{Li2024_TunableCouplerPulseCalibration}. Gate-bias relaxation measurements, leakage diagnostics~\cite{Wood2018_LeakageRB}, and refined pulse calibration~\cite{Li2024_TunableCouplerPulseCalibration,Zhang2026_CZCalibrationWorkflow} provide practical tools for targeting and evaluating these improvements.

\section{Spectral Design and Scalability}
\label{sec:tct_scalability}

Figure~\ref{fig:tct_scalability}(a) sketches a possible nearest-neighbor lattice of transmon data qubits connected by C-shunt couplers $C_{ij}$. The computational qubits remain transmons, for which established control and readout methods are available~\cite{Koch2007_Transmon,Spring2025_intrinsicPCfilterRO,YulinWu2021_Zhuchongzhi,Arute2019_Sycamore}. We use numerical simulations to examine where a coupler two-excitation collision lies relative to a selected gate-control band and how activating one gate changes the residual ZZ interactions on the inactive links of a six-qubit chain.

For the representative cell $Q_1$--$C_{12}$--$Q_2$, the bare states $|020\rangle$ and $|101\rangle$, with occupations ordered as Qubit~1, coupler, Qubit~2, become degenerate at $\omega_c^{\mathrm{col}}=\bar\omega_q-\alpha_c/2$, where $\bar\omega_q=(\omega_1+\omega_2)/2$. This condition follows from the uncoupled energies in Eq.~\eqref{eqn:tft_hamiltonian}. Positive coupler anharmonicity shifts this collision below the fixed \SI{5.55}{GHz}--\SI{6.10}{GHz} gate-control band [Fig.~\ref{fig:tct_scalability}(b)]. For comparison, we use a transmon--transmon--transmon (TTT) coupler anharmonicity of \SI{-0.370}{GHz} reported by Xu \emph{et al.}~\cite{YuanXu2020_TTTCZgate} and a model value of \SI{-0.140}{GHz} used by Stehlik \emph{et al.}~\cite{Stehlik2021_TTT}. Within our fixed frequency allocation, these values place the collision inside the band or only \SI{10}{MHz} below it. In contrast, the device-model value $\alpha_c/2\pi=\SI{0.703}{GHz}$ gives a minimum clearance of \SI{431.5}{MHz}, increasing to \SI{580}{MHz} for an assumed \SI{1}{GHz} anharmonicity. These numerical results indicate that positive coupler anharmonicity can alleviate the frequency constraint imposed by this two-excitation collision.

To assess crosstalk on inactive links, we simulate the six-qubit chain shown in Fig.~\ref{fig:tct_scalability}(a). The chain follows the path $Q_1$--\allowbreak$Q_2$--\allowbreak$Q_3$--\allowbreak$Q_6$--\allowbreak$Q_5$--\allowbreak$Q_4$ and includes its five couplers. All other array links are omitted. Only the frequency of $C_{12}$ is shifted from \SI{6.630}{GHz} to \SI{5.860}{GHz}, while the spectator couplers remain at the idle bias with their interactions retained. For each link $e$, we define $\Delta\zeta_{ZZ}^{(e)}=\zeta_{ZZ}^{(e)}(\mathrm{gate})-\zeta_{ZZ}^{(e)}(\mathrm{idle})$. Figure~\ref{fig:tct_scalability}(c) plots $|\Delta\zeta_{ZZ}^{(e)}|/2\pi$, which is \SI{3.97}{MHz} on the target link, \SI{9.18}{kHz} on the adjacent link (0.231\% of the target response), \SI{20.3}{Hz} at distance 2, and below \SI{1}{Hz} at distances 3 and 4. The response is therefore concentrated near the activated cell. Studies of coupler-bias and frequency optimization in related architectures~\cite{Valles2026_FrequencyPositioning,Heunisch2023_CSFQcplrtheory} motivate further control of spectator interactions in this setting.

The separation of the collision from the gate-control band and the localization of interaction changes near the active gate provide useful design guidance for larger TCT networks. Frequency-collision mitigation in fixed-frequency transmon arrays~\cite{Hertzberg2021_freqcollision} and frequency optimization with tunable couplers~\cite{Valles2026_FrequencyPositioning} offer useful strategies for accommodating parameter variations in larger networks.

\section{\label{sec:discussion_conclusion} CONCLUSION AND DISCUSSION}

We have experimentally demonstrated a TCT device in which a three-junction C-shunt flux-qubit coupler mediates a CZ gate between two transmon qubits. The measured $\zeta_{ZZ}/2\pi$ changes sign over a range from approximately \SI{-5.2}{MHz} to \SI{10.5}{MHz}, providing an idle point with $\zeta_{ZZ}\simeq0$ and strongly suppressed exchange. At this bias, simultaneous single-qubit RB yields $X$-gate fidelities of 99.711(32)\% for Qubit~1 and 99.826(56)\% for Qubit~2, consistent with the individual values. A \SI{45}{ns} flux pulse produces the required conditional phase and implements a nonadiabatic CZ gate with an interleaved-RB fidelity estimate of 99.23(15)\%.

For the measured data-qubit lifetimes and a fixed optimized pulse, relaxation-only simulations predict that increasing the coupler lifetime from approximately \SI{1.1}{\micro s} to \SI{5}{\micro s} raises the CZ fidelity from 99.68\% to nearly 99.8\%. Future work can combine improved coupler coherence with mitigation of dephasing, leakage, flux-pulse distortion, and residual phase errors~\cite{Sung2021_TTT,Li2024_TunableCouplerPulseCalibration,Zhang2026_CZCalibrationWorkflow,Negirneac2021_CZ}.

The numerical analysis shows that positive coupler anharmonicity separates the $|020\rangle$--$|101\rangle$ collision from the selected gate-control band, and gate-induced changes in $\zeta_{ZZ}$ decrease rapidly with distance in the six-qubit chain. These results guide frequency allocation and spectator-interaction control in larger TCT networks.

\section*{ACKNOWLEDGEMENTS}

This work was supported by the National Natural Science Foundation of China (Grants Nos.~U25A6009, T2121001, T2322030, 12475017, 92365301, 92565301, 12574540, 92265207, and 12204528), the Quantum Science and Technology--National Science and Technology Major Project (Grant No.~2021ZD0301800), the Beijing Nova Program (Grant No.~20220484121), and the Ministry of Science and Technology project (Grant No.~2025YFE0217600).

\appendix

\section{Wiring Layout}
\label{appendix:measurement_setup}

\begin{figure}[t]
    \centering
    \includegraphics[width=0.95\linewidth]{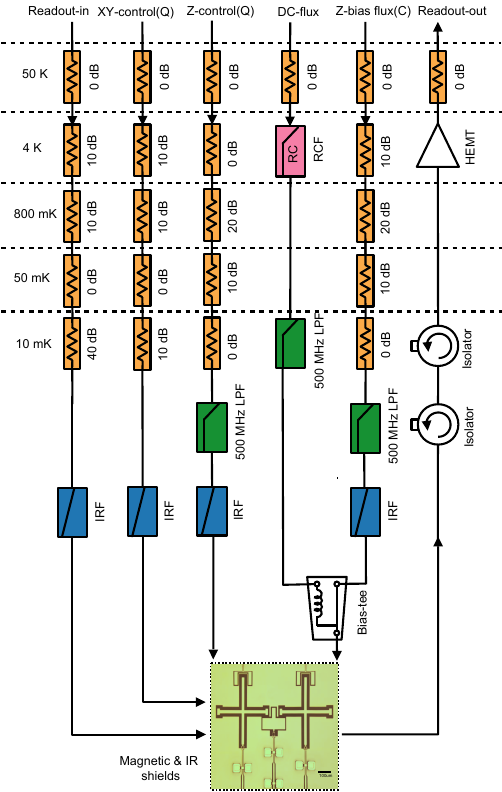}
    \caption{RF, DC-flux, flux-pulse, and readout wiring used for device measurement. The input lines include staged attenuation and infrared filtering. The C-shunt flux line combines DC and pulsed control through a bias tee, and the output chain uses isolators and low-temperature and room-temperature amplification.}
    \label{fig:cryogenic_setup}
\end{figure}

The measurement setup is shown in Fig.~\ref{fig:cryogenic_setup}. The columns show the readout input, qubit XY and Z controls, coupler DC-flux and flux-pulse controls, and readout output. From top to bottom, the diagram follows the cooling stages toward the \SI{10}{mK} device stage in a BlueFors XLD-1000 dilution refrigerator. The coaxial lines include staged attenuation and filtering to suppress thermal noise and infrared radiation before the signals reach the device.

At room temperature, the high-frequency XY signals and low-frequency Z biases are combined and sent to the qubits for state excitation and frequency modulation, respectively. The XY microwave control signals are generated by arbitrary waveform generators based on RFSoC direct digital synthesis (DDS), with the tones synthesized directly in the second Nyquist zone of high-sampling-rate digital-to-analog converters. This removes the conventional IQ-mixing chain, simplifying the control hardware and avoiding additional mixing noise. On the coupler line, only the low-frequency Z bias is delivered. To extend the available bias range and reduce distortion of fast-Z pulses, the output of the DC source and the pulsed Z waveform are combined through bias tees before reaching the coupler.

The readout tone is generated by mixing the AWG waveform with the LO signal through an IQ mixer and is sent to the readout input line. The signal collected from the readout output line is first amplified at low temperature and then further amplified at room temperature. The RF signal is then down-converted to an intermediate frequency and digitized by analog-to-digital converters (ADCs).

\section{Fabrication}
\label{appendix:fabrication}

The device was fabricated using standard micro- and nanofabrication techniques. The capacitors, readout resonators, and transmission lines were fabricated following the procedures described in our previous work on C-shunt flux qubits~\cite{Zhao2026_CshuntHighFidelityGates} and transmon qubits~\cite{Li2025_CoherentErgotropy}.

A distinctive feature of the present process is the simultaneous fabrication of the transmon and C-shunt Josephson junctions. To simplify the process flow, all junction areas were defined in a single lithography step, and the Al/AlO$_x$/Al junctions were then formed in one double-angle evaporation sequence. This shared process requires careful calibration of the junction areas using preliminary test structures. In particular, the small-to-large junction-area ratio sets $\alpha_J$ in the three-junction coupler loop, while the relative junction areas of the transmons and coupler determine their Josephson energies and hence the hybridized level structure.

Table~\ref{tab:tct_design} compares the design values of the selected circuit parameters with the experimental values extracted from spectroscopy. The qubit capacitance $C_1$ is obtained by fitting the qubit spectroscopy data, whereas $C_{\mathrm{sh}}$, $I_c$, and $\alpha_J$ are obtained by fitting the coupler spectrum using the model described in Appendix~\ref{app:tct_device_spectroscopy}. This comparison provides a quantitative check of the device fabrication against the design targets.

\begin{table}[!ht]
    \centering
    \caption{Design and spectrum-derived circuit parameters of the TCT device. The junction-area ratio $\alpha_J$ is distinct from the mode anharmonicities $\alpha_i$.}
    \label{tab:tct_design}
    \small
    \renewcommand{\arraystretch}{1.15}
    \setlength{\tabcolsep}{3pt}
    \begin{tabular*}{\columnwidth}{@{\extracolsep{\fill}}p{0.57\columnwidth}cc@{}}
        \hline
        \hline
        Parameter (unit) & \shortstack{Design\\value} & \shortstack{Measured\\value} \\
        \hline
        \raggedright Qubit 1 capacitance, $C_1$ (fF) & $9$ & $8.335$ \\
        \raggedright Shunt capacitance, $C_{\mathrm{sh}}$ (fF) & $40$ & $36.666$ \\
        \raggedright Large-junction critical current,\newline $I_c$ (nA) & $128.1$ & $118.04$ \\
        \raggedright Junction-area ratio, $\alpha_J$ & $0.435$ & $0.443$ \\
        \hline
        \hline
    \end{tabular*}
\end{table}

\section{Circuit Hamiltonian Derivation for the TCT Device}
\label{app:tct_circuit_hamiltonian}

\begin{figure}[t]
    \centering
    \includegraphics[width=\linewidth,trim=120 110 120 90,clip]{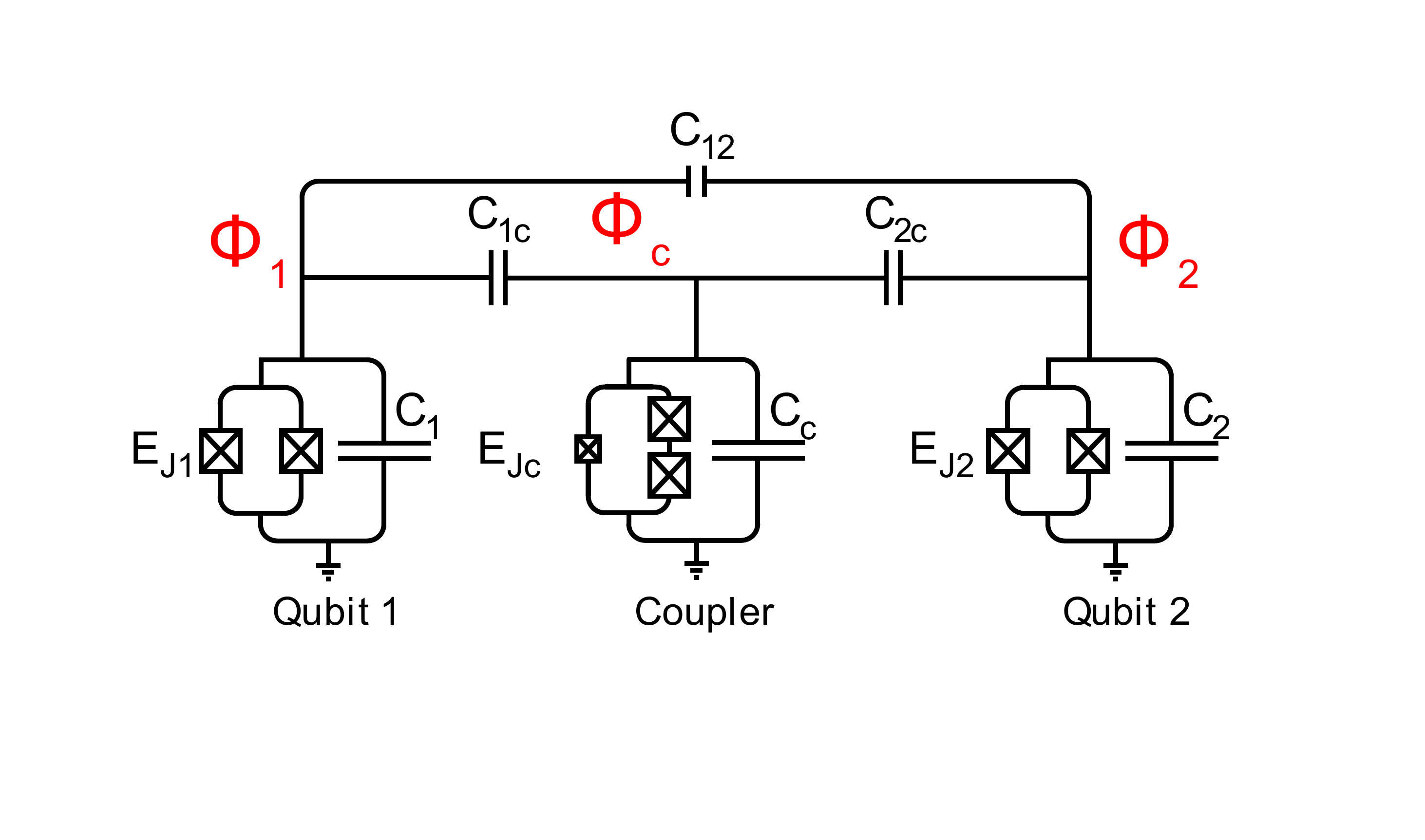}
    \caption{Equivalent circuit of the TCT device, with node fluxes $\Phi_1$, $\Phi_c$, and $\Phi_2$. The data-qubit SQUIDs have effective Josephson energies $E_{J1}$ and $E_{J2}$. The coupler contains two junctions of energy $E_{Jc}$ and a smaller junction of energy $\alpha_J E_{Jc}$. The node capacitances include junction contributions. $C_{1c}$ and $C_{2c}$ couple the qubits to the coupler, while $C_{12}$ connects the qubits directly.}
    \label{fig:tct_circuit_appb}
\end{figure}

Following C-shunt circuit models~\cite{Yan2016_CshuntFluxQubit,Heunisch2023_CSFQcplrtheory} and the capacitive-coupler treatment of the TTT architecture~\cite{FeiYan2018_TTTtheory}, we derive the three-mode Hamiltonian from \fref{fig:tct_circuit_appb}. The key difference from TTT is the three-junction coupler potential, which supports positive anharmonicity.

\subsection{Circuit Lagrangian and quantization}

Define $\theta_\mu=2\pi\Phi_\mu/\Phi_0$ for $\mu=1,c,2$ and $f=2\pi\Phi_{\mathrm{ext}}/\Phi_0$, where the applied coupler flux $\Phi_{\mathrm{ext}}$ is distinct from the dynamical node flux $\Phi_c$. We neglect loop inductance and treat the flux biases as fixed parameters. The data-qubit potentials are $-E_{Ji}\cos\theta_i$, with the SQUID phase offsets absorbed into the coordinate origins.

For the coupler, flux quantization gives $\delta_a+\delta_b-\theta_c=f$. With $\delta_{a,b}=(\theta_c+f)/2\pm\chi$, the loop potential is
\begin{align}
 U_c(\theta_c,\chi;f)={}&-2E_{Jc}\cos[(\theta_c+f)/2]\cos\chi\notag\\
 &-\alpha_J E_{Jc}\cos\theta_c.
 \label{eqn:tct_two_coordinate_potential}
\end{align}
For nearly identical large junctions and a sufficiently large shunt capacitance, we assume that the internal mode $\chi$ lies well above the retained coupler transitions. In the reduced model used here, we set $\chi=0$ near the operating well and neglect its zero-point renormalization. The resulting node capacitance is $C_c=C_{\mathrm{sh}}+(\alpha_J+1/2)C_J$, before adding stray capacitance, where $C_J$ is the capacitance of each large junction. Thus the reduced potential is
\begin{equation}
 U=-\sum_{i=1,2}E_{Ji}\cos\theta_i+U_c(\theta_c,0;f).
\end{equation}

The capacitors in \fref{fig:tct_circuit_appb} give
\begin{align}
 \mathcal L_C={}&\frac12\sum_\mu C_\mu\dot\Phi_\mu^2
 +\frac12\sum_{i=1,2}C_{ic}(\dot\Phi_i-\dot\Phi_c)^2\notag\\
 &+\frac{C_{12}}2(\dot\Phi_1-\dot\Phi_2)^2
 =\frac12\dot{\boldsymbol\Phi}^{T}\mathbf C\dot{\boldsymbol\Phi},
 \label{eqn:tct_lagrangian_capacitance_appb}
\end{align}
where $\boldsymbol\Phi^T=(\Phi_1,\Phi_c,\Phi_2)$ and
\begin{equation}
 \mathbf C=\begin{pmatrix}
 C_{\Sigma1}&-C_{1c}&-C_{12}\\
 -C_{1c}&C_{\Sigma c}&-C_{2c}\\
 -C_{12}&-C_{2c}&C_{\Sigma2}
 \end{pmatrix}.
 \label{eqn:tct_capacitance_matrix}
\end{equation}
Here $C_{\Sigma i}=C_i+C_{ic}+C_{12}$ and $C_{\Sigma c}=C_c+C_{1c}+C_{2c}$. From $\mathcal L=\mathcal L_C-U$, the conjugate charges are $\mathbf Q=\mathbf C\dot{\boldsymbol\Phi}$, giving $H=\mathbf Q^T\mathbf C^{-1}\mathbf Q/2+U$. With $\hat n_\mu=\hat Q_\mu/(2e)$ and $[\hat\theta_\mu,\hat n_\nu]=i\delta_{\mu\nu}$,
\begin{align}
 \hat H={}&\sum_\mu4E_{C\mu}\hat n_\mu^2+U(\hat{\boldsymbol\theta};f)
 +\sum_{\mu<\nu}J_{\mu\nu}\hat n_\mu\hat n_\nu,\label{eqn:tct_charge_hamiltonian}\\
 E_{C\mu}&=\frac{e^2}{2}(\mathbf C^{-1})_{\mu\mu},\qquad
 J_{\mu\nu}=4e^2(\mathbf C^{-1})_{\mu\nu}.
 \label{eqn:tct_charging_parameters}
\end{align}
For $C_{12}\ll C_{ic}\ll C_\mu$, $E_{C\mu}\simeq e^2/(2C_\mu)$ and
\begin{equation}
 J_{ic}\simeq\frac{4e^2C_{ic}}{C_iC_c},\qquad
 J_{12}\simeq\frac{4e^2}{C_1C_2}\left(C_{12}+\frac{C_{1c}C_{2c}}{C_c}\right).
 \label{eqn:tct_capacitive_weak_limit}
\end{equation}
The second term in $J_{12}$ arises from the inverse capacitance matrix and is distinct from the virtual exchange derived below.

\subsection{C-shunt spectrum and three-mode Hamiltonian}

For the average large-junction phase $\varphi_c=(\theta_c+f)/2$, the conjugate charge is $n_{\varphi c}=2n_c$. The local coupler Hamiltonian becomes
\begin{align}
 \hat H_c={}&4E_{Cc}^{(\varphi)}\hat n_{\varphi c}^{\,2}
 -E_{Jc}\left[2\cos\hat\varphi_c
 +\alpha_J\cos(2\hat\varphi_c-f)\right],\notag\\
 E_{Cc}^{(\varphi)}={}&E_{Cc}/4.
 \label{eqn:tct_reduced_cshunt_hamiltonian}
\end{align}
At $f=\pi$, expansion around $\varphi_c=0$ gives
\begin{align}
 U_c={}&U_0+E_{Jc}(1-2\alpha_J)\varphi_c^2\notag\\
 &+\frac{E_{Jc}(8\alpha_J-1)}{12}\varphi_c^4+O(\varphi_c^6).
 \label{eqn:tct_cshunt_quartic}
\end{align}
The quadratic and quartic coefficients are both positive for $1/8<\alpha_J<1/2$. When the quartic term is perturbative, the leading anharmonicity is
\begin{equation}
 \alpha_c\simeq\frac{E_{Cc}^{(\varphi)}}{\hbar}
 \frac{8\alpha_J-1}{1-2\alpha_J},
 \label{eqn:tct_positive_anharmonicity}
\end{equation}
in contrast to the transmon result $\alpha_i\simeq-E_{Ci}/\hbar$. This expansion fails as $\alpha_J\to1/2$, where the quadratic stiffness vanishes. For quantitative calculations, including flux biases away from half flux, diagonalization of Eq.~\eqref{eqn:tct_reduced_cshunt_hamiltonian} gives
\begin{equation}
 \omega_c=\frac{\epsilon_1-\epsilon_0}{\hbar},\qquad
 \alpha_c=\frac{\epsilon_2-2\epsilon_1+\epsilon_0}{\hbar}.
\end{equation}

Projecting the charge interactions onto the local eigenstates gives the lowest-transition coupling
\begin{equation}
 g_{\mu\nu}=\frac{J_{\mu\nu}}{\hbar}
 |\langle0_\mu|\hat n_\mu|1_\mu\rangle
   \langle0_\nu|\hat n_\nu|1_\nu\rangle|.
 \label{eqn:tct_charge_to_g}
\end{equation}
Choosing $\hat n_\mu\simeq i|n_{01}^{(\mu)}|(\hat a_\mu^\dagger-\hat a_\mu)$ and approximating the local spectra by Duffing ladders yields
\begin{align}
 \hat H/\hbar={}&\sum_{\mu=1,c,2}\left[
 \omega_\mu\hat a_\mu^\dagger\hat a_\mu
 +\frac{\alpha_\mu}{2}\hat a_\mu^\dagger\hat a_\mu^\dagger\hat a_\mu\hat a_\mu\right]\notag\\
 &+\sum_{i=1,2}g_{ic}(\hat a_i^\dagger\hat a_c+\hat a_i\hat a_c^\dagger)\notag\\
 &+g_{12}(\hat a_1^\dagger\hat a_2+\hat a_1\hat a_2^\dagger)\notag\\
 &-\sum_{i=1,2}g_{ic}(\hat a_i^\dagger\hat a_c^\dagger+\hat a_i\hat a_c)\notag\\
 &-g_{12}(\hat a_1^\dagger\hat a_2^\dagger+\hat a_1\hat a_2),
 \label{eqn:tct_full_hamiltonian_appb}
\end{align}
where $\hat a_\mu$ and $\hat a_\mu^\dagger$ are the same mode operators as in Eq.~\eqref{eqn:tft_hamiltonian}. The last two lines contain the counter-rotating pair-creation and pair-annihilation terms. Omitting these terms reproduces Eq.~\eqref{eqn:tft_hamiltonian}. For a strongly anharmonic coupler, quantitative multilevel calculations should retain the actual charge matrix elements rather than assume oscillator scaling.

\subsection{Effective exchange and residual ZZ interactions}

Let $\Delta_{ic}=\omega_i-\omega_c$, $\Sigma_{ic}=\omega_i+\omega_c$, and $\Delta_{12}=\omega_1-\omega_2$. In the dispersive regime $g_{ic}\ll|\Delta_{ic}|$, the Schrieffer--Wolff transformation $\hat H'=e^{\hat S}\hat H e^{-\hat S}$ uses
\begin{align}
 \hat S=\sum_{i=1,2}\Big[&\frac{g_{ic}}{\Delta_{ic}}
 (\hat a_i^\dagger\hat a_c-\hat a_c^\dagger\hat a_i)\notag\\
 &-\frac{g_{ic}}{\Sigma_{ic}}
 (\hat a_i^\dagger\hat a_c^\dagger-\hat a_c\hat a_i)\Big].
 \label{eqn:tct_sw_generator_appb}
\end{align}
With $\hat H_{\mathrm h}/\hbar=\sum_\mu\omega_\mu\hat a_\mu^\dagger\hat a_\mu$ and $\hat V$ denoting the qubit-coupler interaction terms in Eq.~\eqref{eqn:tct_full_hamiltonian_appb}, the generator satisfies $[\hat S,\hat H_{\mathrm h}]=-\hat V$. The second-order term $[\hat S,\hat V]/2$, projected onto the coupler ground state, gives~\cite{FeiYan2018_TTTtheory}
\begin{align}
 g_{\mathrm{eff}}={}&g_{12}+\frac{g_{1c}g_{2c}}2
 \left(\frac1{\Delta_{1c}}+\frac1{\Delta_{2c}}
 -\frac1{\Sigma_{1c}}-\frac1{\Sigma_{2c}}\right).
 \label{eqn:tct_geff_appb}
\end{align}
This is the leading single-excitation exchange within the three-mode approximation. The harmonic generator does not eliminate all nonlinear multilevel couplings. The $1/\Sigma_{ic}$ terms are counter-rotating corrections. For $\omega_c>\omega_{1,2}$, the mediated exchange is negative and can cancel the positive bare coupling $g_{12}$.

The residual ZZ interaction strength is defined by the dressed energies,
\begin{equation}
 \zeta_{ZZ}=\frac{E_{101}-E_{100}-E_{001}+E_{000}}{\hbar},
 \label{eqn:tct_zz_definition_appb}
\end{equation}
with states ordered as Qubit~1, coupler, Qubit~2. The relevant double-excitation gaps are
\begin{align}
 d_1&=-\Delta_{12}-\alpha_1,\qquad d_2=\Delta_{12}-\alpha_2,\notag\\
 d_c&=\Delta_{1c}+\Delta_{2c}-\alpha_c,
\end{align}
defined as the bare energy of $|101\rangle$ minus that of $|200\rangle$, $|002\rangle$, and $|020\rangle$, respectively, divided by $\hbar$. Retaining the leading number-conserving terms of the perturbative expansion in Ref.~\cite{FeiYan2021_TTTCZgatetheory} gives
\begin{align}
 \zeta_{ZZ}^{(2)}={}&2g_{12}^2\left(\frac1{d_1}+\frac1{d_2}\right),
 \label{eqn:tct_zz_second_order_appb}\\
 \zeta_{ZZ}^{(3)}={}&4g_{12}g_{1c}g_{2c}\left[
 \frac1{d_1\Delta_{2c}}+\frac1{d_2\Delta_{1c}}
 +\frac1{\Delta_{1c}\Delta_{2c}}\right],
 \label{eqn:tct_zz_third_order_appb}
\end{align}
\begin{align}
 \zeta_{ZZ}^{(4)}={}&2g_{1c}^2g_{2c}^2\Bigg[
 \frac1{d_c}\left(\frac1{\Delta_{1c}}+\frac1{\Delta_{2c}}\right)^2\notag\\
 &\hspace{4em}+\frac1{d_1\Delta_{2c}^2}+\frac1{d_2\Delta_{1c}^2}\Bigg],
 \label{eqn:tct_zz_fourth_order_appb}\\
 \zeta_{ZZ}\simeq{}&\zeta_{ZZ}^{(2)}+\zeta_{ZZ}^{(3)}+\zeta_{ZZ}^{(4)}.
 \label{eqn:tct_zz_final_appb}
\end{align}
The second-order term comes from direct transitions to $|200\rangle$ and $|002\rangle$, while the third-order term describes interference between direct and coupler-mediated paths. At fourth order, the two paths through $|110\rangle$ and $|011\rangle$ reach $|020\rangle$ with amplitude $\sqrt2g_{1c}g_{2c}(1/\Delta_{1c}+1/\Delta_{2c})$, producing the $1/d_c$ term. The expressions include the subtraction of the single-excitation energy shifts and vanish in the harmonic limit.

This truncation retains terms proportional to $g_{12}^2$, $g_{12}g_{1c}g_{2c}$, and $g_{1c}^2g_{2c}^2$, as appropriate near exchange cancellation, where $g_{12}$ is of order $g_{1c}g_{2c}/|\Delta_{ic}|$. Higher mixed terms and counter-rotating corrections to $\zeta_{ZZ}$ are omitted. When states involved in a virtual transition approach resonance, numerical diagonalization is required. Unlike exchange cancellation, cancellation of $\zeta_{ZZ}$ also depends on $\alpha_c$ through $d_c$. The positive C-shunt anharmonicity therefore changes the balance of double-excitation channels relative to TTT, allowing the two idle conditions to be optimized together.

\section{TCT Device Spectroscopy and Coupler-Spectrum Fit}
\label{app:tct_device_spectroscopy}

The flux-dependent spectra shown in \fref{fig:zz_vs_cplrflux}(b,c) were measured by biasing the C-shunt coupler with a square $Z$ pulse while applying an XY probe pulse to one data qubit. The probe frequency was swept to map the coupler-dependent transitions, and the crosstalk response on the other data qubit provided an additional spectroscopy channel. The pulse-amplitude axis was converted to the reduced flux coordinate $\Phi_c/\Phi_0$ using the calibrated linear flux conversion. The spectroscopy response is plotted as a normalized transmission amplitude in \fref{fig:zz_vs_cplrflux}(b,c), while the overlaid curves represent dressed transitions of the three-mode Hamiltonian of the TCT device.

Within the frozen high-frequency-mode approximation of Appendix~\ref{app:tct_circuit_hamiltonian}, the bare C-shunt branch is modeled as
\begin{align}
    H_c/h =&\,
    4(E_{Cc}^{(\varphi)}/h)\hat{n}_{\varphi c}^{\,2}
    -(E_{Jc}/h)
    \big[
    2\cos\hat{\varphi}_{c}
    \notag \\
    &+
    \alpha_J\cos\left(2\hat{\varphi}_{c}
    -2\pi\Phi_c/\Phi_0\right)
    \big],
    \label{eqn:cshunt_fit_hamiltonian_appc}
\end{align}
where $\varphi_c$ is the average large-junction phase, $n_{\varphi c}$ is its conjugate charge, and $E_{Cc}^{(\varphi)}=E_{Cc}/4$ as derived in Eq.~\eqref{eqn:tct_reduced_cshunt_hamiltonian}. Here the spectroscopy coordinate $\Phi_c$ denotes the externally applied coupler flux (the $\Phi_{\mathrm{ext}}$ of Appendix~\ref{app:tct_circuit_hamiltonian}), rather than the dynamical node flux in Fig.~\ref{fig:tct_circuit_appb}. An irrelevant constant offset has been omitted, and $\alpha_J$ is the small-to-large junction-area ratio. Numerical diagonalization of Eq.~\eqref{eqn:cshunt_fit_hamiltonian_appc} gives the flux-dependent transition frequency and anharmonicity,
\begin{align}
    f_c(\Phi_c)=\frac{\epsilon_1(\Phi_c)-\epsilon_0(\Phi_c)}{h},
    \notag \\
    \frac{\alpha_c(\Phi_c)}{2\pi}=
    \frac{\epsilon_2(\Phi_c)-2\epsilon_1(\Phi_c)+\epsilon_0(\Phi_c)}{h}.
    \label{eqn:cshunt_fit_transition_appc}
\end{align}
Near the coupler-frequency minimum, a possible local parametrization is
\begin{equation}
    f_c(\Phi_c)=
    \sqrt{
    f_{c,\min}^{2}
    +v_c^{2}
    \left(
    \Phi_c/\Phi_0-\Phi_s/\Phi_0
    \right)^{2}},
    \label{eqn:cshunt_local_dispersion_appc}
\end{equation}
where $f_{c,\min}$ is the minimum coupler transition frequency, $\Phi_s$ is the flux bias at the coupler sweet spot, and $v_c$ has units of frequency because the flux displacement is dimensionless. For the full scan, the numerically calculated spectrum can be used directly in place of this local approximation.

The avoided crossings with Qubit~1 and Qubit~2 are described by projecting Eq.~\eqref{eqn:tft_hamiltonian} onto the one-excitation manifold in the basis $\{|100\rangle,|001\rangle,|010\rangle\}$, with state labels ordered as Qubit~1, coupler, Qubit~2:
\begin{equation}
    H_{1}/h=
    \begin{pmatrix}
        f_{1} & g_{12}/2\pi & g_{1c}(\Phi_c)/2\pi \\
        g_{12}/2\pi & f_{2} & g_{2c}(\Phi_c)/2\pi \\
        g_{1c}(\Phi_c)/2\pi & g_{2c}(\Phi_c)/2\pi & f_{c}(\Phi_c)
    \end{pmatrix},
    \label{eqn:tct_spectroscopy_fit_hamiltonian}
\end{equation}
Here $f_i=\omega_i/2\pi$ are bare mode frequencies, the $g$ parameters are coupling strengths expressed in angular-frequency units, and the vacuum energy is set to zero. The qubit-coupler matrix elements generally depend on flux. In the harmonic approximation, $g_{ic}(\Phi_c)\simeq g_{ic}(\Phi_{\mathrm{ref}})\sqrt{f_c(\Phi_c)/f_c(\Phi_{\mathrm{ref}})}$, with a specified reference bias $\Phi_{\mathrm{ref}}$. Treating them as constants is a local approximation. The transition curves are the eigenvalues
\begin{equation}
    f_m^{\mathrm{fit}}(\Phi_c)
    =
    \lambda_m\left[
    H_{1}(\Phi_c)/h
    \right],
    \qquad m=1,2,3,
    \label{eqn:tct_spectroscopy_eigenvalue_fit_appc}
\end{equation}
which are fitted to the resolved branches in \fref{fig:zz_vs_cplrflux}(b,c). The broad spectrum constrains the flux-dependent coupler dispersion, while the enlarged spectrum resolves the avoided crossings that constrain the qubit frequencies and coupling strengths. To extract the coupler anharmonicity, we evaluate $f_c(\Phi_c)$ from the circuit Hamiltonian in Eq.~\eqref{eqn:cshunt_fit_hamiltonian_appc}, rather than from the local frequency parametrization alone. The spectroscopy fit yields $g_{1c}/2\pi\simeq\SI{101}{MHz}$, $g_{2c}/2\pi\simeq\SI{105}{MHz}$, and $g_{12}/2\pi\simeq\SI{7.5}{MHz}$. With the fitted circuit parameters, the three lowest coupler eigenenergies determine the anharmonicity through Eq.~\eqref{eqn:cshunt_fit_transition_appc}, giving the model-derived value $\alpha_c/2\pi\simeq\SI{703}{MHz}$ reported in Table~\ref{tab:tft_param}.

\FloatBarrier

\bibliographystyle{apsrev4-2}
\bibliography{references}

\end{document}